\documentclass[aoas,MSNbibl,nameyear,dvips]{arximspdf}

\doi{10.1214/10-AOAS398J}
\referstodoi{10.1214/10-AOAS398}
\volume{5}
\issue{1}
\pubyear{2011}
\firstpage{91}
\lastpage{95}

\begin{document}
\begin{frontmatter}

\title{Discussion of: A statistical analysis of multiple
temperature proxies: Are reconstructions of surface temperatures over the
last 1000~years~reliable?\thanksref{T1}}
\runtitle{Discussion}
\pdftitle{Discussion on A statistical analysis of multiple temperature proxies:
Are reconstructions of surface temperatures over the last 1000 years reliable?
by B. B. McShane and A. J. Wyner}

\thankstext{T1}{We gratefully acknowledge the support of the Weather
and Climate Impacts Assessment Science Program at the National Center
for Atmospheric Research. The National Center for Atmospheric Research
is sponsored by the National Science Foundation.}

\begin{aug}
\author[A]{\fnms{Eugene R.} \snm{Wahl}\corref{}\ead[label=e1]{Eugene.R.Wahl@noaa.gov}%
\ead[label=u1,url]{http://www.ncdc.noaa.gov/paleo/paleo.html}}
\and
\author[B]{\fnms{Caspar M.} \snm{Ammann}\ead[label=e2]{ammann@ucar.edu}%
\ead[label=u2,url,text={\vtop{\hsize=122pt\baselineskip=10pt http://www.cgd.ucar.edu/ccr/ammann/ millennium/MBH-reevaluation.html}}]
{http://www.cgd.ucar.edu/ccr/ammann/millennium/MBH-reevaluation.html}}

\runauthor{E. R. Wahl  and C. M. Ammann}

\affiliation{NOAA's National Climatic Data Center and the~National~Center~for~Atmospheric~Research}

\address[A]{NOAA/NESDIS/NCDC/\\
Paleoclimatology Branch\\
325 Broadway Street\\
Boulder, Colorado 80305\\
USA\\
\printead{e1}\\
\printead{u1}} %adresu isvedimo komanda gale!

\address[B]{National Center\\
for Atmospheric Research\\
1850 Table Mesa Drive\\
Boulder, Colorado 80305\\
USA\\
\printead{e2}\\
\printead{u2}}
\end{aug}

% HISTORY:
\received{\smonth{9} \syear{2010}}
\revised{\smonth{9} \syear{2010}}

% ABSTRACT

% KEYWORDS

\end{frontmatter}

McShane and Wyner (\citeyear{McShane2010}) (MW) introduce new methods into the effort to
reconstruct the large-scale Northern Hemisphere temperature average over
the past millennium, helping to advance interaction between applied
statisticians and paleoclimatologists to improve understanding of
pre-instrumental climates. However, despite a good effort to capture the
various points of contention in the reconstruction arena, MW provide an
incomplete, and at times inadequate, review of the existing literature
considering reconstruction of Northern Hemisphere surface temperatures
over the past millennium. In particular, the evaluations cited regarding
the original Mann, Bradley and Hughes (MBH) reconstruction (\citeyear{Mann1998}/\citeyear{Mann1999})
and MBH's use of principal component (PC) summaries of dendroclimatic
proxy data fail to address this issue properly, and in the process
propagate errors that have been fully addressed in the literature.
Similarly, MW omit important information in their examination of the
methodology outlined and used by Ammann and Wahl (\citeyear{Ammann2007}) (AW) to test the
significance of the MBH reconstruction. Because examinations related to
the MBH reconstruction have had particular salience not only in the
specialist literature, but also politically [cf. House Committee on
Energy and Commerce (\citeyear{HCEC2005}); American Association for the Advancement of
Science (\citeyear{AAAS2005}); Wegman, Scott and Said (\citeyear{Wegman2006}); Russell et al. (\citeyear{Russell2010})] and
in terms of scientific review and advisement [National Research Council
(\citeyear{NRC2006})], it is important that these omissions be corrected. We will
focus our discussion on them, along with providing more general closing
observations.

MBH used PC summaries of tree ring proxy information in different parts
of the world to reduce the weighting of these regions in terms of the
number of proxy data series from them employed in the MBH
reconstruction. The use of PCs (more generally, the use of
eigenvector/singular value decomposition methods) for the purpose of
dimension reduction of data is a common\vadjust{\goodbreak} practice in
climatology/paleoclimatology. The bulk of attention about the use of PC
summarization by MBH has focused on a network  of 70 tree ring data
series from North America used in their reconstruction in the first half of the 15th
century (called ITRDB, from the International Tree Ring Data Base), and
is associated with significant confusion and erroneous claims about the
extent to which the way MBH employed this practice biased their results.
MW add to this confusion.

First, it should be noted that MBH formed the underlying proxy data into
standardized anomalies relative to their calibration period, 1902--1980,
rather than centering and standardizing over the full length of the
data. This was done because the MBH reconstruction consists of 12
segments, each of which was recalibrated to include increasing numbers
of proxy data series as time comes closer to the present. Their practice
ensured that all the proxy data \textit{and} the instrumental data used
in calibration were processed in common terms. This was a reasonable
judgment given the segmental nature of the MBH reconstruction, although
the normal methodology for PC extraction from standardized anomaly data
is to center and standardize over the full data period.

In AW (Supplement, Section~2;
\href{http://www.cgd.ucar.edu/ccr/ammann/millennium/AW\_supplement.html}
{http://www.cgd.ucar.edu/ccr/ammann/}
\href{http://www.cgd.ucar.edu/ccr/ammann/millennium/AW\_supplement.html}
{millennium/AW\_supplement.html}) we
evaluate the impact these two methods of data treatment have on the
extraction of PCs. In the MBH ``common centered'' method, the first PC
contains a noticeable ``hockey stick''-like shape, whereas in the ``full
centered'' method this shape is spread across the first two PCs. When a
vector sum of the first two PCs from both methods is calculated, the
plots of the resultant time series have essentially identical shapes,
with a larger amplitude in the case of the ``common centered'' method. A
point that is generally not appreciated in this context [e.g., National
Research Council (\citeyear{NRC2006}); MW], is that MBH used the first \textit{two}
North American ITRDB PCs in their reconstruction of the time period
1400--1449, which is the key target of critical focus in this regard
[cf. McIntyre and McKitrick (\citeyear{McIntyre2003,McIntyre2005a,McIntyre2005b,McIntyre2005c})] (MM). In Wahl and
Ammann (\citeyear{Wahl2007}) (WA), we systematically examined the difference the use of
the first two PCs from each method actually has in the MBH
reconstruction for this time period, and note that the reconstructed
time series' structures are nearly identical, with a slight average
warming of 0.05 deg. C when ``common centered'' PCs are used (WA, Figure~3, blue range; cf. page~51).

When nonstandardized anomalies are input into a PC algorithm using the
variance--covariance matrix rather than the correlation matrix (as MM;
cf. WA, page~45), the ``hockey-stick''-like shape occurs lower in the PC
order, here PC4. This occurs because the algorithm captures information
in the first one or two eigenvectors/PCs that is primarily related to
the absolute magnitude of the largest-scaled variables in the data,
which here differ by a factor of 13 (AW, Supplement, Sections~2, 4). In
WA, we show that the impact of using only the first two or three ITRDB
PCs extracted this way is a different reconstruction (pink curve, Figure~3), which does not pass validation testing. When the\vadjust{\goodbreak} fourth or fifth
ITRDB PCs are added, the reconstruction converges to those using two PCs
derived from fully standardized data, all of which pass validation (WA,
Figure~3, blue range; cf. AW, Supplement, Section 1).

Concerning the significance testing method outlined and used by AW, we
would like to re-emphasize the consideration that using the full AR
structure of proxy data to drive creation of ``pseudo-proxies'' as null
models for establishing reconstruction skill significance thresholds is
likely overly conservative. More importantly, MW omit from their
examination of our reasoning that we report application of this (in our
view) overly conservative method in pseudo-proxy generation (AW, pages~77--78). Thus, the kind of test MW emphasize, when applied with
real-world context to our emulation of the MBH reconstruction, shows
highly successful validation results. Ten of the 12 segments are
significant at the 95\% level (the other two are significant at the 89\%
and 94\% levels), and the much-discussed 1400--1449 and 1450--1499
segments are significant at 99\% and 96\%, respectively. We wish to
emphasize this point, as it distinctly counters MW's argument concerning
this issue.

Finally, there is an extensive literature contradicting MW's assertions
about low or poor relationships between proxies and climate. The climate
system certainly exhibits spatio-temporal complexity. Yet
global/hemispheric average temperature is far from being
``nonpredictable.'' With indications about the evolution of a few
primary modes of variability and independent knowledge of key radiative
forcing histories, one can explain the mean temperature evolution,
including its spatial expression, over the time period of instrumental
data [Tett et al. (\citeyear{Tett1999}); Ammann et al. (\citeyear{Ammann2003});
Hansen et al. (\citeyear{Hansen2007})]. Such relationships should hold as well for the last
millennium, which is well illustrated in Figure 6.14 of Jansen et al.
(\citeyear{Jansen2007}); note that the figure demonstrates this result is robust to
uncertainty concerning the strength of the solar irradiance forcing
data. The challenge in reconstructions remains the proper
implementation of the available geophysical knowledge (cf. AW),
including the cross-dependence of its uncertainties. When random samples
systematically outperform predictors known to contain useful
information, such a result suggests the need to reevaluate the
reconstruction model itself (e.g., LASSO). Apparently, important aspects
were missed or improperly implemented. Otherwise, independent proxy
evaluations would long ago have drawn the general conclusion that there
is no skill in capturing spatial and temporal climate, which is not
borne out by the literature.

% imsref loaded by smiklovaite, 2011-01-20 07:59:23
% imsref loaded by smiklovaite, 2011-01-20 08:03:48

\printaddresses

\end{document}